\documentclass[useAMS,usenatbib]{mn2e}
\newif\ifAMStwofonts

\usepackage{amsmath,times}

\newcommand{\lc}{{\scshape LensClean}}
\newcommand{\zd}{z_{\rmn{d}}}
\newcommand{\dd}{d_{\rmn{d}}}
\newcommand{\ds}{d_{\rmn{s}}}
\newcommand{\dds}{d_{\rmn{ds}}}
\newcommand{\vc}[1]{\bmath{#1}}
\newcommand{\rtext}[1]{\quad\text{#1}}
\newcommand{\sub}[1]{_{\mathrm{#1}}}
\newcommand{\hunits}{km$\,$s$^{-1}$$\,$Mpc$^{-1}$}

\title[The Hubble Constant from CLASS B0218+357 using the Advanced Camera for Surveys]
{The Hubble Constant from gravitational lens CLASS B0218+357 using the Advanced Camera for Surveys}

\author[]{T. York$^{1}$, N. Jackson$^{1}$, I.W.A. Browne$^{1}$, O.
Wucknitz$^{2}$, J.E. Skelton$^{1\dagger}$\\
$^{1}$University of Manchester, Jodrell Bank Observatory, Macclesfield, 
Cheshire, SK11 9DL \\
$^{2}$Universit\"{a}t Potsdam,   Institut f\"{u}r Physik, Am Neuen
Palais 10, 14469 Potsdam, Germany \\
$\dagger$Current address: Institute for Astronomy, University of Edinburgh, 
Royal Observatory, Blackford Hill, EH9 3HJ }

\pagerange{\pageref{firstpage}--\pageref{lastpage}} \pubyear{2004}
\input psfig.tex
\begin{document}
\Large

\maketitle

\label{firstpage}

\begin{abstract}
We present deep optical observations of the gravitational lens system 
CLASS B0218+357 from which we derive an estimate for the Hubble Constant ($H_0$).
Extensive radio observations using the VLA, MERLIN, the VLBA and VLBI have 
reduced the degeneracies between $H_0$ and the mass model parameters in this lens
to one involving only the position of the radio-quiet lensing galaxy with
respect to the lensed images. B0218+357 has an image separation of only 
334~mas, so optical observations have, up until now, been unable to resolve
the lens galaxy from the bright lensed images. Using the new Advanced Camera
for Surveys, installed on the {\it Hubble Space Telescope} in 2002, we have
obtained deep optical images of the lens system and surrounding field. These
observations have allowed us to determine the separation between the lens galaxy
centre and the brightest image, and so estimate $H_0$. 

We find an optical galaxy position -- and hence an $H_0$ value -- that varies 
depending on our approach to the spiral arms in B0218+357. If the most prominent 
spiral arms are left unmasked, we find $H_0$ = 70$\pm$5~\hunits (95\% confidence). 
If the spiral arms are masked out we find $H_0$ = 61$\pm$7~\hunits (95\% confidence). 

\end{abstract}


\section{Introduction}

Objects at cosmological redshifts may be multiply imaged by the action
of the gravitational field of foreground galaxies. The first such example of
gravitational lensing was the system 0957+561 (Walsh, Carswell \&
Weymann 1979) in which the core of a background quasar is split into
two images 6\arcsec~apart. Since then approximately 70 cases of
gravitational lensing by galaxies have been found.\footnote{A full
compilation of known galaxy-mass lens systems is given on the CASTLeS 
website at http://cfa-www.harvard.edu/glensdata}

Refsdal (1964) pointed out that such multiple-image gravitational lens 
systems could be used to measure the Hubble constant, if the background source was
variable, by measuring time delays between variations of the lensed
image and inferring the difference in path lengths between the
corresponding ray paths. The combination of typical deflection angles
of $\sim$1\arcsec~around galaxy-mass lens systems with typical
cosmological distances implies time delays of the order of weeks,
which are in principle readily measurable. Time delays have been
measured for eleven gravitational lenses to date: CLASS~B0218+357 (Biggs
et al. 1999; Cohen et al. 2000), RXJ~0911.4+0551 (Hjorth et al. 2002),
0957+561 (Kundic et al. 1997; Oscoz et al. 2001), PG~1115+080 (Schechter et al. 1997)
CLASS~B1422+231 (Patnaik \& Narasimha 2001), SBS~1520+530 (Burud et
al. 2002a), CLASS~B1600+434 (Koopmans et al. 2000; Burud et al. 2002b),
CLASS~B1608+656 (Fassnacht et al. 1999; Fassnacht et al. 2002),
PKS~1830$-$211 (Lovell et al. 1998), HE~2149$-$2745 (Burud et
al. 2002b) and HE~1104$-$1805 (Ofek \& Maoz 2003).  In principle, given a suitable variable source, the
accuracy of the time delay obtained can be better than 5\%. This
has already been achieved in some cases (eg. 0957+561) and there is no doubt that diligent 
future campaigns will further improve accuracy and also produce 
time delays for more gravitational lens systems.

Gravitational lenses provide an excellent prospect of a one-step
determination of $H_0$ on cosmological scales. The major problem is
that, in addition to the time delay, a mass model for the lensing
galaxy is required in order to determine the shape of the
gravitational potential. The model is needed to convert the time delays 
into angular diameter distances for the lens and source. In double-image lens
systems in which the individual images are unresolved this is a serious
problem as the number of constraints on
the mass model (lensed image positions and fluxes) allows no degrees
of freedom after the most basic parameters characteristic of the
system (source position and flux together with galaxy mass,
ellipticity and position angle) have been fitted. In four-image
systems the extra constraints provide assistance, and in a few cases,
such as the ten-image lens system CLASS B1933+503 (Sykes et al. 1998)
more detailed constraints on the galaxy mass model are exploited (Cohn
et al. 2001).

There are two further systematic and potentially very serious
problems.  The first is that the radial mass profile of the lens is
almost completely degenerate with the time delay, and hence $H_0$
(Gorenstein, Shapiro \& Falco 1988; Witt, Mao \& Keeton 2000; Kochanek
2002). Given a time delay, $H_0$ scales as $2-\beta$, where
$\beta$ is the profile index of the potential, $\phi \propto r^\beta$.
Work by Koopmans \& Treu (2003) shows that mass profiles may vary from an isothermal
slope by up to 10\% for single galaxies, producing corresponding uncertainties in $H_0$. 
The problem is particularly serious for four-image systems, because 
the images are all approximately the same distance from the centre of the lens and
thus constrain the radial profile of the lensing potential poorly. On
the other hand, for CLASS B1933+503, with three sources producing ten 
images, the radial mass profile is well constrained (Cohn et al. 2001). 
Unfortunately B1933+503 does not show radio variability (Biggs et al. 2000) 
and is optically so faint that measuring a time delay is likely to be very hard.

In some cases Einstein rings may provide
enough constraints, despite the necessity to model the extended source
which produces them (Kochanek, Keeton \& McLeod 2001) although models
constrained by rings may still be degenerate in $H_0$ (Saha \&
Williams 2001).

The mass profile degeneracy is particularly sharply illustrated by
``non-parametric'' modelling of lens galaxies (Williams \& Saha 2000;
Saha \& Williams 2001). Such models assume only basic physical
constraints on the galaxy mass profile, such as a monotonic decrease
in surface density with radius. They find consistency with the
observed image data for a wide range of galaxy mass models, which are
themselves consistent with a wide range of $H_0$. Combining two
well-constrained cases of lenses with a measured time delay, CLASS
B1608+656 and PG~1115+080, Williams \& Saha (2000) find
H$_0=61\pm18$\hunits (90\% confidence). 

There are a number of approaches to the resolution of the mass profile degeneracy
problem. One is to assume that galaxies have approximately isothermal mass distributions.
($\beta\sim1$). There are two parts to the lensing argument in favour of isothermal galaxies: 
from the lack of odd images near the centre of observed lens
systems Rusin \& Ma (2001) are able to reject the hypothesis that
significant number of lensing galaxies have profiles which are much
shallower ($\beta>1.2$) than isothermal, assuming a single power-law model
is appropriate for the mass contained interior to the lensed images. Similarly it can be
shown that models which are significantly steeper than isothermal are
unable to reproduce constraints from positions and fluxes in
existing lenses with large numbers of constraints (e.g. Mu\~noz,
Kochanek \& Keeton 2001; Cohn et al. 2001). The most straightforward
approach, that of assuming an isothermal lens, has been taken by many
authors. In most cases this yields $H_0$ estimates of between 55 and
70~\hunits\ (e.g. Biggs et al. 1999; Koopmans \& Fassnacht 1999;
Koopmans et al. 2000; Fassnacht et al. 1999) but studies of some
lenses imply much lower values (e.g. Schechter et al. 1997; Barkana
1997; Kochanek 2003). In fact, Kochanek (2003) finds a serious
discrepancy with the {\it HST} key project value of $H_0$=71\hunits\ (Mould
et al.  2000; Freedman et al. 2001) unless, far from being isothermal,
galaxy mass profiles follow the light distribution. 

Falco, Gorenstein \& Shapiro (1985) pointed out the second important
problem. Any nearby cluster produces a contribution to the lensing
potential in the form of a convergence which is highly degenerate with
the overall scale of the lensing system and hence with $H_0$.
Unfortunately, the systems with the most accurately determined time
delays and the best-known galaxy positions are often those with large angular
separation such as 0957+561, and these are the systems in which
lensing is most likely to be assisted by a cluster. Again progress can
be made by appropriate modelling of the cluster, and many attempts
have been made to do this for 0957+561 (e.g. Kundic et
al. 1997; Bernstein \& Fischer 1999; Barkana et al. 1999) although
there remain uncertainties in the final $H_0$ estimate. As
an alternative, the optical/infra-red images of the host galaxy may make an important
contribution towards the breaking of degeneracies (Keeton et al. 2000).

Kochanek \& Schechter (2004) summarise the contribution of lensing
to the $H_0$ debate so far and present options for further
progress. One approach is simply to accumulate more $H_0$
determinations and rely on statistical arguments to iron out the
peculiarities which affect each individual lens system; this approach
is vulnerable only to a systematically incorrect understanding of galaxy
mass profiles. The alternative approach is to select a lens system
in which additional observational effort is most capable of decreasing
the systematic errors on the $H_0$ estimate to acceptable levels. In this
paper we argue that CLASS B0218+357 is the best candidate for this
process. We describe new {\it Hubble Space Telescope} ({\it HST}) observations
using the Advanced Camera for Surveys (ACS) which are aimed at
removing the last major source of systematic uncertainty in this
system. We then describe how we use the imaging data to
derive a value for the Hubble constant.

\section{CLASS B0218+357 as a key object in $H_0$ determination}

CLASS B0218+357 was discovered during the early phase of the CLASS
survey, the Jodrell Bank-VLA Astrometric Survey (JVAS; Patnaik et al. 1992). 
B0218+357 consists of two images (A and B) of a background flat-spectrum radio source separated by 0\farcs334,
together with an Einstein ring (Patnaik et al. 1993). The optical
spectrum shows a red continuum source superimposed on a galaxy
spectrum. The redshift of the lensing galaxy has been measured
optically by Browne et al. (1993) and Stickel \& Kuehr (1993), and 
by Carilli, Rupen \& Yanny (1993) at radio wavelengths giving
the most accurate result of 0.6847. Cohen et al. (2003) have 
measured the source redshift of 0.944.

It quickly became apparent that the lensing was performed by a 
spiral galaxy. Spiral lenses generally produce smaller image 
separations than elliptical lenses because of their lower mass.
The spiral nature of B0218+357 was deduced directly from early high-resolution optical images from the Nordic
Optical Telescope (Grundahl \& Hjorth 1995), and was consistent with evidence from molecular line
studies which revealed absorption of the radio emission from the
background radio source by species in the lensing galaxy including 
CO, HCO$^{+}$, HCN (Wiklind \& Combes 1995), formaldehyde (Menten \& 
Reid 1996) and water (Combes \& Wiklind 1997). Moreover, in the optical,
the A image, which is further from the galaxy, is fainter than the B 
image (Grundahl \& Hjorth 1995) , despite being a factor $\sim$3 
brighter in the radio. This suggests that the line of sight to the A 
image intercepts a great deal of dust, possibly associated with a giant 
molecular cloud in the galaxy. The lensing galaxy appears close to face-on, a 
conclusion deduced from its symmetrical appearance in optical images. 
This conclusion is consistent with the small velocity line-width of the absorption lines (e.g. 
Wiklind \& Combes 1995).

Further radio imaging resolved both the A and B images into core-jet structures 
(Patnaik, Porcas \& Browne 1995; Kemball et al. 2001; Biggs et al. 2003), as well as revealing more 
details of the Einstein ring (Biggs et al. 2001). The combined constraints from the core-jet
structure and the ring together strongly constrain mass models. A and B lie at different
distances from the galaxy, and with the Einstein ring constraints permits determination of 
both the angular structure of the lensing mass (Wucknitz et al. 2004) and (most importantly) the
mass--radius relation for the lens.

Biggs et al. (1999) have determined a time delay of 10.5$\pm$0.4~days
(95\% confidence) for
B0218+357 using radio monitoring observations made with the VLA at
both 8.4~GHz and 15~GHz. Consistent results were obtained from the
variations in the total intensity, the percentage polarization and
the polarization position angle. Biggs et al. used the time delay and
existing lens model to deduce a value for the Hubble constant of 
69$^{+13}_{-19}$ \hunits (95\% confidence). It should be noted, however, 
that the error bars on the assumed position for the lensing galaxy with 
respect to the lensed images were over-optimistic and hence their quoted 
error on $H_0$ is too small. Cohen et al. (2000) also observed
B0218+357 with the VLA, and measured a value for the time delay of 
10.1$^{+1.5}_{-1.6}$ days. This corresponds to an $H_0$ value of 71$^{+17}_{-21}$ \hunits 
(95\% confidence), the larger error bars in Cohen et al.'s measurement being due 
to their use of a more general model for the source variability, 
although they used the same model for the lensing effect as Biggs et al. 
The error bars do not take into account any systematic error associated with 
the uncertain galaxy position. 

Leh\'{a}r et al. (2000) used the then existing constraints to model the CLASS
B0218+357 system. They found that, even for isothermal models, the
implied value of $H_0$ was degenerate with the position of the centre of
the lensing galaxy, with a change of about 0.7 \hunits (about 1~per~cent)
in the value of $H_0$ for every 1~mas shift in the central galaxy position. 
Their uncertainty on the position derived from {\it HST} infra-red observations
using the NICMOS camera is approximately $\pm$30~mas.

Recently, using a modified version of the \lc\ algorithm (Kochanek \& Narayan 1992; 
Ellithorpe et al. 1996; Wucknitz 2004), Wucknitz et al. (2004) have been able
to constrain the lens position from radio data of the Einstein ring.
With the time delay from Biggs et al. (1999) of $(10.5\pm0.4)$~days, they
obtain for isothermal models a value of H$_0=78\pm6$~\hunits (95\% confidence). They use VLBI results from other authors
(Patnaik et al. 1995, Kemball et al. 2001) as well as their own data 
(Biggs et al. 2003) to constrain the radial profile from the image 
substructure and obtain $\beta\approx1.04$, very close to isothermal ($\beta=1$).
Our aim in this paper is to use new optical observations to determine the lensing galaxy
position directly and to compare this with the indirect determination of
Wucknitz et al. (2004).

We briefly summarise the reasons why, given the observations presented
here, CLASS B0218+357 offers the prospect of the most unbiased and
accurate estimate of $H_0$ to date.

\begin{enumerate}
\item The observational constraints are arguably the best available
for any lens system with a measured time-delay.

\item The radio source is relatively bright (a few hundred mJy at GHz
frequencies) and variable at radio frequencies, so time delay monitoring
is relatively straightforward and gives a small error (Biggs et al. 1999)
which can be improved with future observations.

\item The system is at relatively low redshift. This means that the
derived values for $H_0$ will not depend on the matter density parameter
and cosmological constant by more than a few percent.

\item The lens is an isolated single galaxy and there are no field
galaxies nearby to contribute to the lensing potential (Leh\'{a}r et al.,
2000).

\end{enumerate}

Although most lenses have at least one of these desirable properties,
CLASS~B0218+357 is the only one known so far which has all of them.
It thus becomes a key object for $H_0$ determination. It is only the
lack of an accurate galaxy position that led in the past to it being excluded
from consideration by many authors (e.g. Schechter 2001; Kochanek 2003).

\section{The ACS observations}

Resolving the lens galaxy and lensed images is an aim that benefits 
from high resolution combined with high dynamic range, and so we
asked for and were awarded time on the Advanced Camera for Surveys 
(ACS; Clampin et al. 2000) on the {\it Hubble Space Telescope}. 

Although observations in the blue end of 
the spectrum would have maximised angular resolution, the likely morphological type 
of the lensing galaxy (Sa/Sab) meant that asymmetry due to star formation in 
the spiral arms could have caused problems in the deconvolution process which
relies on symmetry in the lensing galaxy (see Section 5).
Hence observations at wavelengths longer than 4000\AA~in the rest frame of 
the galaxy were desirable.  At the redshift of the lens this dictated 
the use of {\it I} band, i.e. the F814W filter on the {\it HST}.

\begin{figure}
\psfig{figure=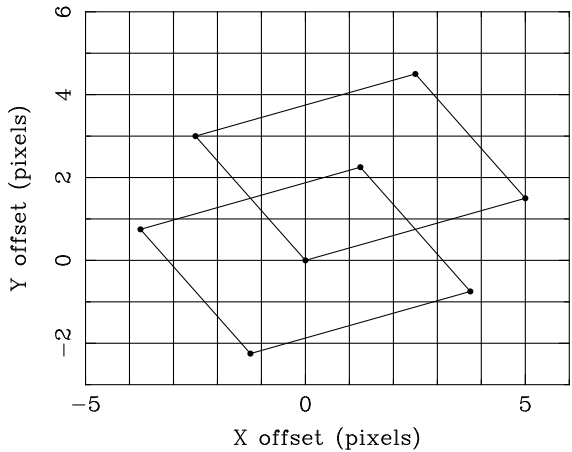,width=7.5cm,angle=-90}

\caption{\label{DithPat}4+4 dither pattern used for the observations of
0218+357. The pattern provides steps to the level of $\frac{1}{4}$-pixel.}
\end{figure}

The ACS has two optical/near-IR ``channels'', the Wide Field Channel (WFC) 
and the High Resolution Channel (HRC). The HRC's pixel response exhibits 
a diffuse halo at longer wavelengths due to scattering within the CCD. 
As a result, roughly 10\% of the flux from a point source 
will be scattered into this halo at 8000\AA, possibly making it more difficult to
resolve the lensing galaxy from the lensed images. We selected the WFC 
for use in our observing programme since it does not suffer from this effect. 
Unlike the HRC, the WFC moderately under-samples the {\it HST} point-spread function (PSF) 
at 8000\AA. To counter this effect, we selected two distinct four-point dither patterns,
alternating between them over the course of the observations. The dither 
patterns used are shown in Figure \ref{DithPat} (see also Mutchler \& Cox 2001
for more information on {\it HST} dither patterns). 


The WFC has a field of view of 202\arcsec$\times$202\arcsec, and 
a plate scale of approximately 50~mas pixel$^{-1}$. 
Since we used a gain of unity, saturation in the images occurs near the 16-bit 
analogue to digital conversion limit of 65,000~e$^-$ pixel$^{-1}$ rather than 
at the WFC's full well point of 85,000~e$^-$ pixel$^{-1}$. We determined the
exposure time required on B0218+357 through simulation.


The full programme of B0218+357 observations was carried out over the period 
from August 2002 to March 2003. Details of the observing dates and exposure 
times are shown in Table \ref{ACSObsLog}. 
The total available telescope time was split into 7 visits on B0218+357,
6 of which provided 2 hours integration time on the science target. The 
remaining visit (16) was designed to permit the programme to be salvaged 
in the unlikely case that the observing pattern chosen for visits 10-15 
proved to be both inappropriate and uncorrectable. This visit suffered
from increased observing overheads relative to the other visits and 
provided an integration time of 1 hour, 22 minutes on B0218+357 as a result.
In order to deconvolve the images we required an ACS/WFC PSF, so two short
visits (1 and 2 in Table \ref{ACSObsLog}) were dedicated to observing two 
Landolt standard stars (Landolt 1992). Following McLure et al. (1999), 
observations were taken with several different exposure times to allow the 
construction of a composite PSF that would have good signal-to-noise in 
both core and wings whilst avoiding saturation of the core. Standards were 
selected to be faint enough not to saturate the WFC chip on short integration
times, and to have the same $V-I$ colour, to within 0.2 magnitudes, as
the lensed images in the B0218+357 system. The exposure times on the standard
stars ranged from 0.5 seconds to 100 seconds each, the longest exposures each being
split into two 50 second exposures to simplify cosmic ray rejection.

\begin{table*}
\begin{tabular}{lllrcc}
Visit no.&Target&Observation date & Exposure time & Dither pattern& File name root\\
&&&&&\\
10 & CLASS B0218+357 & 2003 February 28   & 20$\times$360 sec & 4+4 & j8d410 \\
11 & CLASS B0218+357 & 2003 March 01      & 20$\times$360 sec & 4+4 & j8d411 \\
12 & CLASS B0218+357 & 2003 January 17-18 & 20$\times$360 sec & 4+4 & j8d412 \\
13 & CLASS B0218+357 & 2003 March 06      & 20$\times$360 sec & 4+4 & j8d413 \\
15 & CLASS B0218+357 & 2003 March 11      & 20$\times$360 sec & 4+4 & j8d415 \\
14 & CLASS B0218+357 & 2002 October 26-27 & 20$\times$360 sec & 4+4 & j8d414 \\
16 & CLASS B0218+357 & 2002 September 11  & 12$\times$360 sec & 4+4 & j8d416 \\
&                    &                    &  8$\times$75 sec & 4C  &    \\
 1 & 92 245	     & 2002 October 17-18 &  8$\times$0.5 sec  & 4+4 & j8d401 \\
&                    &                    &  8$\times$8 sec & 4+4 & \\
&                    &                    &  1$\times$360 sec& -   & \\
 2 & PG0231+051B     & 2002 August 25     &  8$\times$0.5 sec   & 4+4 & j8d402 \\
&                    &                    &  8$\times$8 sec & 4+4 & \\
&                    &                    &  8$\times$50 sec& 4C  & \\
\end{tabular}
\caption{\label{ACSObsLog}
Log of {\it HST} observations. All observations were taken with the ACS 
using the F814W filter, corresponding to {\it I} band. A 4+4 dither pattern refers 
to an eight-point dither consisting of two nested parallelograms, whereas 4C refers to a
four-point dither parallelogram with the exposure at each point split
into two for explicit cosmic-ray rejection. }
\end{table*}

\section{Reduction of the ACS data}

The uncalibrated data produced by automatic processing of raw {\it HST} 
telemetry files by the OPUS pipeline at the Space Telescope Science
Institute (STScI) were retrieved along with flat fields, superdarks 
and other calibration files.

The data were processed through the ACS calibration pipeline, {\it
CALACS} (Pavlovsky et al. 2002), which runs under NOAO's {\sc IRAF} software. The {\it
CALACS} pipeline de-biased, dark-subtracted and flat-fielded the data,
producing a series of calibrated exposures. The pipeline also combined
the CR-SPLIT exposures in visits 1, 2 and 16 to eliminate cosmic
rays. The calibrated exposures were in general of acceptable quality
for use in the next stage of reduction, except for visit 15 in which
there was some contamination of the images by stray light, probably from
a WFPC2 calibration lamp (R. Gilliland and M. Sirianni, private communication). It is
possible that this defect can be corrected in the future using
the techniques which were used by Williams et al. (1996) to remove stray light
from some HDF exposures, but we have not attempted to deconvolve the contaminated visit.

The calibrated exposures were fed to the next stage of reduction,
based around the {\it dither} package (Fruchter \& Hook 2002), and the STSDAS packages
{\it pydrizzle} (Hack 2002) and {\it multidrizzle}
(Koekemoer et al. 2002). These tools clean cosmic rays, remove the ACS
geometric distortion and ``drizzle'' the data on to a common output
image (Fruchter \& Hook 2002). The drizzle method projects the input
images on to a finer grid of output pixels. Flux from each 
input pixel is distributed to output pixels according to the degree of overlap between
the input pixel and each output pixel. To successfully combine dithered images into a single output 
image, knowledge of the pointing offsets between exposures is needed. The expected offsets 
are determined by the dither pattern used, but the true offsets might vary from 
those expected due to thermal effects (Mack et al. 2003) within single visits. 
To determine 
the true pointing offsets between dithered exposures, we cross-correlated the images.
Since the WFC at I-band moderately undersamples the HST's PSF, and since most of the features
detected in the images were extended rather than stellar, we opted for pixel-by-pixel cross-correlation 
rather than comparisons of positions of stars between different pointings, to maximise our use of the 
available information. 

Images in each visit were drizzled on to a common
distortion-corrected frame and then pairs of these images were cross-correlated. The 
two-dimensional cross-correlations have a Gaussian shape near their centres. The shift between pairs of 
images is measured by fitting a Gaussian function to the peak in the cross-correlation, and the 
estimated random error in the shift is derived from the position error given by the Gaussian fit.
For our data, the random errors in the measured shifts ranged from 0.8 to 2.5~mas. The RMS 
scatter between corresponding pointings within a visit was typically less than 10~mas, or 20\% 
of a single WFC pixel. We fed these shifts to the {\it multidrizzle} script,
which carried out the drizzling of visits to common, undistorted output frames.
As part of the drizzling process, we opted to decrease the output pixel 
size from 50~mas square (the natural size of the undistorted output pixels) to 25~mas square. 
To avoid blurring and ``holes'' in the output image, the input pixels were shrunk to 70\% of their 
nominal size before being drizzled on to the output frames (Fruchter \& Hook 2002). We used a gaussian
drizzling kernel to slightly improve resolution and reduce blurring. At the end of this process each visit, 
except for visit 15, provided us with a output image with improved sampling compared to the individual 
input exposures. The deconvolution of these images is described in Section \ref{AnalysisPos}.

Since deconvolution depends greatly on the accuracy of the PSF model, we have produced a number 
of different PSFs. Unfortunately the Landolt standard star (Landolt 92 245) observed in visit 1 was resolved into 
a 0.5\arcsec~double by the ACS/WFC, so we have concentrated on extracting a PSF from visit 2 (Landolt PG0231+051B). 
The calibrated exposures were combined using {\it multidrizzle}. Saturated pixels were masked. The resulting PSF
suffered from serious artifacts consisting of extended wings approximately 80~mas up and down the chip from the 
central peak, possibly due to imperfect removal and combination of pixels which were
affected by bleeding of saturated columns. We believe these extended wings to be artifacts since they
rotate with the telescope rather than the sky, and are not present in stars in other visits. 

In lieu of ideal standard star PSFs, we have created per-visit PSFs by averaging 
field stars together. These field star PSFs do not suffer from the artifacts present in the visit 2 PSF.
Manual examination, pixel-by-pixel, of the difference between the PSFs and the central 
regions close to image B, indicate that the RMS error in the PSFs is between 5 and 15\%. 
Such an error increases linearly with the counts, rather than with their square root; 
lacking a perfectly-fitting PSF, we have allowed for this error when performing the data analysis
described in Section \ref{AnalysisPos}. 

\section{Analysis of the position data and results}
\label{AnalysisPos}

\subsection{General remarks}

\begin{figure}
\psfig{figure=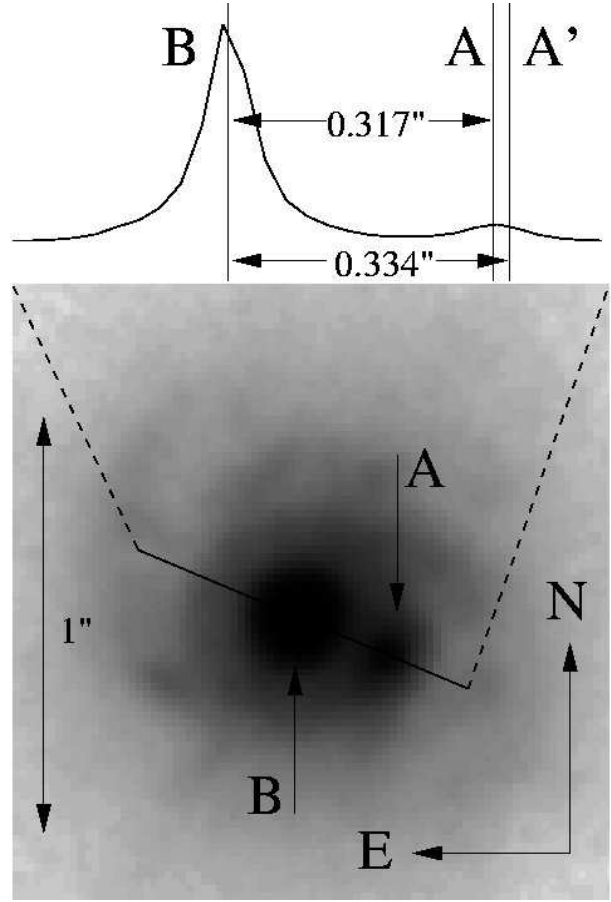,width=8cm}
\caption{\label{AllImg}Combined ACS image of B0218+357. The lensed images are both
visible; the brighter image, B, is close to the centre of the lensing
galaxy. The spiral arms of the galaxy are clearly visible. The plot above the image
shows a one dimensional slice passing through images A and B. The best-fit positions
of A and B on this slice are marked, along with A', the position of A expected
from the radio image separation (334~mas). The separation between A and B in the 
optical image is 317$\pm$4~mas ($2\,\sigma$).}
\end{figure}

Figure \ref{AllImg} shows the ACS image of the CLASS B0218+357 system from
the combined dataset, consisting of all science visits on B0218+357 excluding visit 15. 
To produce this combined image, separate visits were related through restricted linear transformations 
(rotation and translation) based on the positions of unsaturated stars common to all images, and then
co-added. In locating the lensing galaxy, however, we did not use this combined image, preferring to 
work with the separate visits. The two compact images (A and B) can be distinguished as can the lensing galaxy 
which lies close to B. The spiral arms of the galaxy are clearly seen confirming the earlier
deductions that the lensing galaxy is a spiral (Wiklind \& Combes
1995; Carilli et al. 1993). The spiral arms appear to be smooth and regular and there is no sign of significant clumping
associated with large-scale star formation. The galaxy appears almost exactly face-on. We deduce this by 
assuming a galaxy position close to B and comparing counts between pixels at 90 degree angles from each other
about the assumed centre. Examination of the residuals reveals no sign of ellipticity. 

The core of the lensing galaxy is strongly blended with B (Figure
\ref{AllImg}) and is relatively weak. Extrapolation of an exponential
disk fit to the outer isophotes of a slice through the central region
shows that the peak surface brightness of image B exceeds that of the
galaxy by a factor of about 30-50. Thus the determination of an
accurate position for the galaxy is a challenging task. Before
discussing the process in more detail below, we outline the various steps we
go through to obtain a galaxy position. They are:

\begin{enumerate}
\item For each visit we measure the positions of the A and B images.
\item We subtract PSFs from these positions.
\item Using PSF-subtracted data we look for the galaxy position about 
which the residuals left after PSF subtraction appear most symmetric. We do not subtract a galaxy model 
from the images. This approach finds the centre of the most symmetric 
galaxy consistent with the data.
\item We compute the mean and variance of the galaxy positions found from the 
individual visits.
\end{enumerate} 

\subsection{Analysis procedure}

Our reduced ACS/WFC images are over 8450x8500 pixels in size, and cover 202\arcsec x202\arcsec
on the sky. We cut out a region of 128x128 pixels (3.2\arcsec x3.2\arcsec) centred on the lens 
and analyse this to make fitting computationally practical and to isolate the lens from 
other objects on the sky.

Since the images, particularly image B, have much higher surface brightnesses
than the galaxy, their positions can be located relatively accurately by subtracting
parametric or empirical PSFs from the data and minimising the residuals. Fits were
carried out on circular regions centred on the brightest pixel of each compact image. The 
regions chosen were 11 pixels in diameter. To avoid any bias arising from the choice of PSF, 
we used both parametric models (Airy and Gaussian functions) and the field star PSFs in the fits. 
The field star PSFs were consistently better fits to both A and B than the parametric models, 
Gaussians being insufficiently peaked and Airy functions having diffraction
rings that were too prominent. A linear sloping background was modelled along with the PSF 
in order to take account of the flux due to the galaxy. Typically the various methods agreed 
on positions to within a tenth of a pixel (2.5~mas). 

The separation of A and B determined by optical PSF fitting is consistently less than the radio 
separation of 334~mas. We find that the mean image separation in the optical is 317$\pm$2~mas (1$\,\sigma$)
when the field star PSFs are used, and 315$\pm$4~mas when Gaussian PSFs are used. The 
corresponding result for Airy function PSFs is 311$\pm$10~mas. These values are mean separations 
taken over the six processed visits on B0218+357. We have checked the plate scale of drizzled images
against stars listed in the US Naval Observatory's B1.0 catalogue, and find that the nominal drizzled
plate scale of 25~mas\,pixel$^{-1}$ is correct to better than 1\% for all visit images.

An anomalous optical image separation has been 
suggested before for B0218+357, starting with ground-based optical imaging by Grundahl \& Hjorth (1995) 
and again by Hjorth (1997). Jackson, Xanthopoulos \& Browne (2000) used NICMOS imaging to find an image 
separation of 318$\pm$5~mas, in agreement with our result from field star PSFs.

We hypothesise that that this low separation may be a result of
the high, and possibly spatially variable, extinction in the region of
A. We suggest that some of the image A optical emission arises from
the host galaxy rather than from the AGN which dominates the B image
emission. Thus the centroid of A may not be coincident with the AGN
image. Image A may be obscured completely and the emission 
seen could be due to a large region of star formation associated with
an obscuring giant molecular cloud. In view of this possibility and the fact that B is
consistently much brighter than A in the optical, we measure galaxy 
positions as offsets from our measured position for B. Thus, although A's position
may be distorted in the optical it does not directly influence our measurements of $H_0$.

\begin{figure}
\psfig{figure=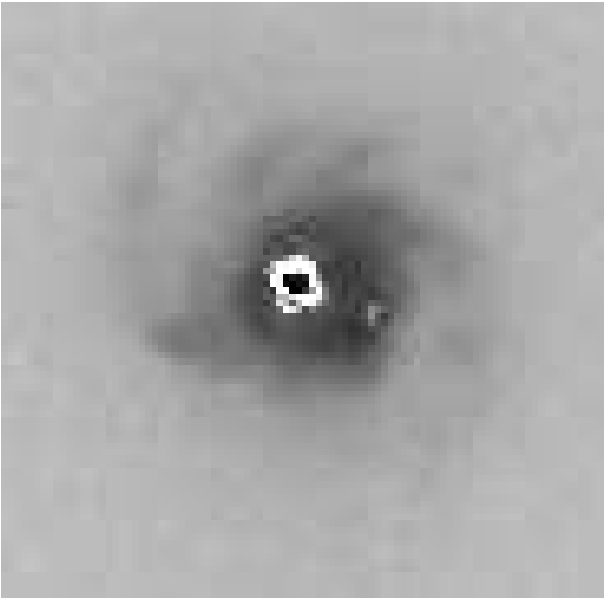,width=8cm}
\caption{\label{Subtracted}A visit image with A and B subtracted after
fitting fluxes and positions. The residuals near the centre of each 
subtracted image show maxima of approximately 20\% of the unsubtracted 
light.}
\end{figure}

Having determined positions and fluxes for A and B, we created an image 
containing two PSFs as a model for the flux from the lensed images alone (the ``model image'').
A model image was made separately for each visit and subtracted from each observed image, 
leaving a residual image which contained only the galaxy plus subtraction errors. 
Figure \ref{Subtracted} shows a typical image with A and B subtracted. The model allowed
us to keep track of how much PSF flux was removed from each pixel in producing the 
residual images. 

We opted to use the criterion of maximum symmetry in the residuals as a
goodness-of-fit parameter rather than attempting to fit a parametric model, 
such as an exponential disk profile, to the light distribution
of the galaxy. The symmetry fit statistic is expressed as

\begin{equation}
\chi^2 = \sum_{P} \frac{{\left| s({\bf r}) - s({\bf r}')\right|}^2}{\sigma({\bf r})^2 + \sigma({\bf r}')^2},
\label{SymFitStat}
\end{equation}

in which $s({\bf r})$ is the count-rate at image pixel position ${\bf r} = (x, y)$ in counts per second, 
$\sigma({\bf r})$ is the estimated noise at ${\bf r}$, 
$P$ is the set of pixels included in the fit (see below) and
${\bf r}'$ is the reflection of ${\bf r}$ around the galaxy position ${\bf g}$, given by simple geometric
considerations as

\begin{equation}
{\bf r}' = 2{\bf g} - {\bf r}.
\label{Reflection}
\end{equation}

The random error for each pixel in the image, $\sigma({\bf r})$ is estimated from the CCD equation 
(Merline \& Howell 1995) summed with a contribution from the assumed random error in the PSF. 
The estimate of the noise is given by

\begin{equation}
\sigma({\bf r})^2 = \frac{q^4}{\tau^2} \frac{q^{-2}\tau s({\bf r}) + S + R^2}{N} + \mu^2 s_p({\bf r})^2,
\label{NoiseEq}
\end{equation}

where $\tau$ is the integration time at a single dithered pointing, 
$\mu$ is the assumed fractional error in the PSF,
$s_p({\bf r})$ is the count-rate from A and B alone (stored in the model image),
$q$ is the drizzling scale factor (the ratio between output and input pixel size, 0.5 for these images)
$N$ is the number of dithered pointings combined in the drizzle process,
$S$ is the sky noise and $R$ is the ACS/WFC read noise. Both sky noise
and read noise are expressed in units of electrons, and the integration time
is in seconds. The resulting noise figure has units of counts per second,
as does the drizzled image. We have checked this noise estimate against the 
background noise in our images and against simulations of the drizzling process.

The set of pixels (P) included in the calculation of this $\chi^2$ figure can
bias the fit if it is ill-chosen. When ${\bf r}'$ falls outside
the boundaries of the image, the pixel ${\bf r}$ is considered to contribute 
nothing to the $\chi^2$ statistic and the pixel is not included in the set P.
Such pixels therefore do not contribute to the number of constraints available
and as a result do not increase the number of degrees of freedom in the fit.
Alternative treatments can introduce bias; for instance, if these pixels 
are assigned large $\chi^2$ values the fitting program is biased towards
placing the galaxy in the geometrical centre of the image. If the same 
pixels are considered to contribute zero towards the $\chi^2$ statistic but are
still counted as part of the set P, they increase the number
of degrees of freedom in the fit and bias the fit to positions away from the 
image's geometrical centre. To avoid these possibilities we do not count degrees 
of freedom from pixels whose reflection about the galaxy centre ends up outside
the image boundaries.

The symmetry criterion is non-parametric and has the 
advantage of minimizing the assumptions that are imposed on the data; 
the use of a particular distribution as a function of radius in any case 
contains an implicit assumption of symmetry. Using the symmetry criterion 
on its own is in principle robust whether or not the galaxy has a central 
bulge, and should also be unaffected if the galaxy contains a bar. The symmetry 
criterion will also hold for galaxies having moderate 
inclinations to the line of sight. For a circularly symmetric galaxy,
the main effect of a small deviation away from a face-on orientation will be 
to render the observed image slightly elliptical. The basic symmetry criterion 
is that points and their reflections about the true centre of the galaxy's image should
have the same flux (to within the measurement errors). Therefore whether the galaxy's 
image has circular or slightly elliptical isophotes is unimportant because in both
cases the same isophote passes through both point and reflection.
This argument breaks down for spirals with significant inclinations as absorption
is likely to become important and destroy any symmetry present in the image.

The centre of maximum symmetry could be displaced by spiral arms if they are not themselves symmetric about
the galaxy centre. To analyse the possible effect on $H_0$ of spiral arms we have made two sets of fits.
In the first set we used all the data and did not apply any masking. In the second set we masked off the 
most obvious spiral arms by using an annular mask centred on image B with an inner radius of 0.375\arcsec and 
an outer radius of 0.875\arcsec. Regions within the inner radius or beyond the outer radius of the mask were
left free to contribute to the symmetry fit. We implemented the symmetry fits so that masked pixels did not 
contribute to either the number of degrees of freedom or to the $\chi^2$ value.

The PSF error ($\mu$ in equation \ref{NoiseEq}) can cause a systematic change in galaxy position when varied between 0.05 and 0.15 (5 to 15\%).
An increase of $\mu$ from 0.05 to 0.15 can increase $H_0$ by up to 10~\hunits. For values above 0.15 the systematic change in galaxy position is 
small compared with the random error. We estimate the PSF error separately for each visit, by taking the range of the highest and lowest residuals
and dividing that range by the peak count-rate of image B. We find values between 0.07 (visit 11) and 0.19 (visit 10) for $\mu$. For the other visits, 
the estimated value for $\mu$ is found to be 0.12.  In the remainder of the paper we do not allow $\mu$ to vary freely but fix it to these estimated values.

\subsection{Extraction of the galaxy position}

\begin{figure*}
\psfig{figure=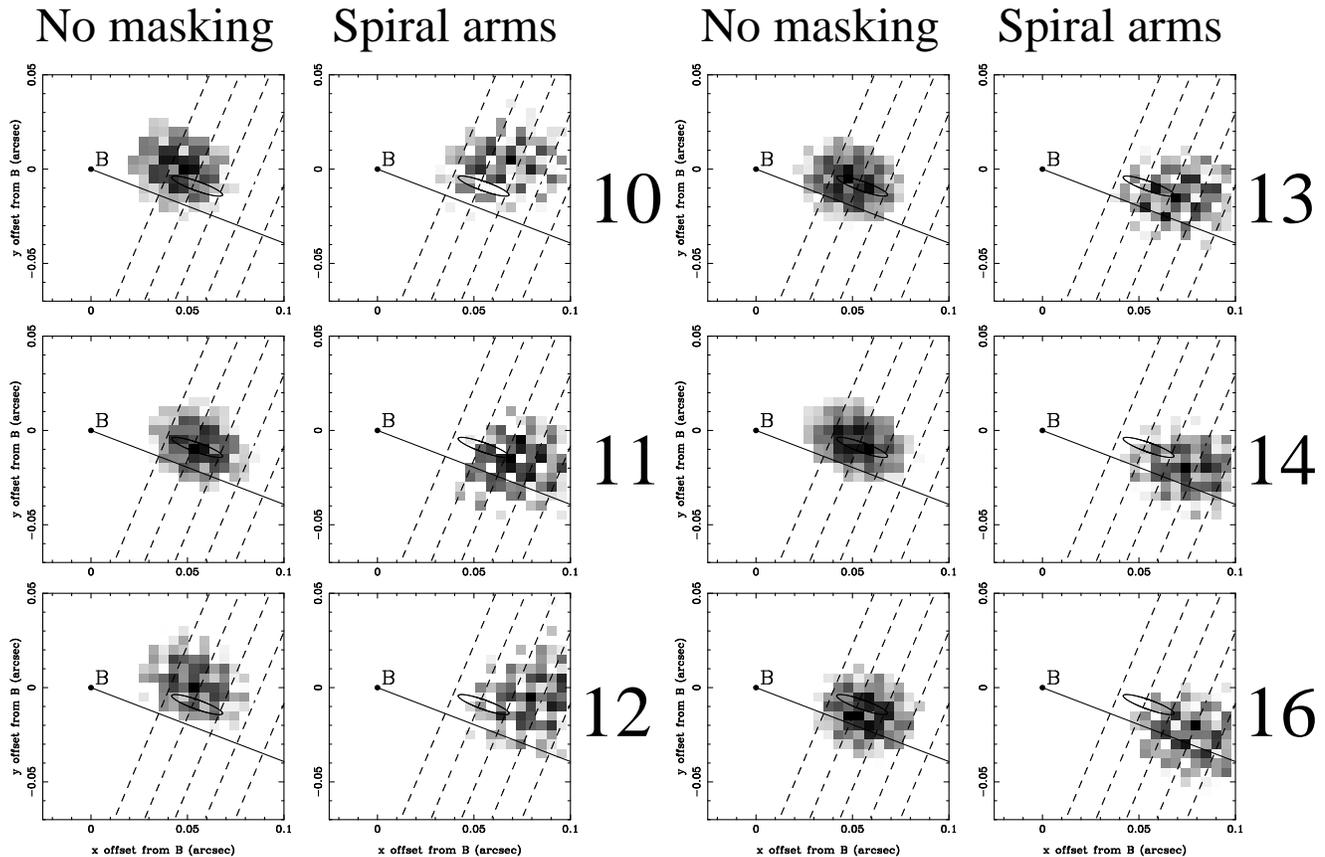,width=17.5cm}
\caption{\label{GridPlot}$\chi^2$ grids for the galaxy position. The visit number is shown to the right of each 
pair of grids. The right-hand plot for each visit shows the effect of masking out the spiral arms, whilst the 
left-hand plot shows the $\chi^2$ grid when no masking is applied. The position of B is marked, as is
a line pointing towards the (radio) A component. The ellipse represents the
position of the galaxy centre found by Wucknitz et al. (2004) using \lc\ modelling of the Einstein ring, and the dotted lines
represent $H_0$ of (90,80,70,60,50)\protect{\hunits} from left to right, assuming
an isothermal model. The axes are RA/Dec offsets from the position of B, expressed in arc-seconds. The RA offset
is given with west as positive.}
\end{figure*}

In applying the symmetry criterion to 0218 we calculated the symmetry $\chi^2$ statistic (i.e. that of equation \ref{SymFitStat})
for a grid of galaxy positions extending 20~mas east to 100~mas west of B, and from 80~mas
south to 50~mas north of B. The spacing between adjacent grid points is 5~mas. We present these 
grids in Figure \ref{GridPlot} for visits 10, 11, 12, 13, 14 and 16. The grids are shown both with and without masking of spiral arms.
Table 2 lists the resulting galaxy positions.

\begin{table}
\begin{tabular}{ccccc}
Visit & \multicolumn{2}{c}{Centre} & \multicolumn{2}{c}{Centre} \\
 & \multicolumn{2}{c}{(No masking)} & \multicolumn{2}{c}{(Spiral arms masked)} \\
 & $\Delta\alpha$ & $\Delta\delta$ & $\Delta\alpha$ & $\Delta\delta$ \\
10 & $+$50 & $+$6 & $+$70 & $+$12 \\
11 & $+$60 & $-$4 & $+$69 & $-$18 \\
12 & $+$59 & $+$9 & $+$84 & $+$8 \\
13 & $+$54 & $-$2 & $+$72 & $-$5 \\
14 & $+$59 & $+$0 & $+$76 & $-$16 \\
16 & $+$61 & $-$6 & $+$79 & $-$14 \\
\\
Mean & $+$57$\pm$4 & $+$1$\pm$6 & $+$75$\pm$6 & $-$6$\pm$13 \\
\end{tabular}

\caption{\label{GridTable}Derived optical centre of the galaxy, expressed as offsets in mas
from the measured optical position of B. RA offsets are given with west as positive.}
\end{table}

\begin{figure}
\psfig{figure=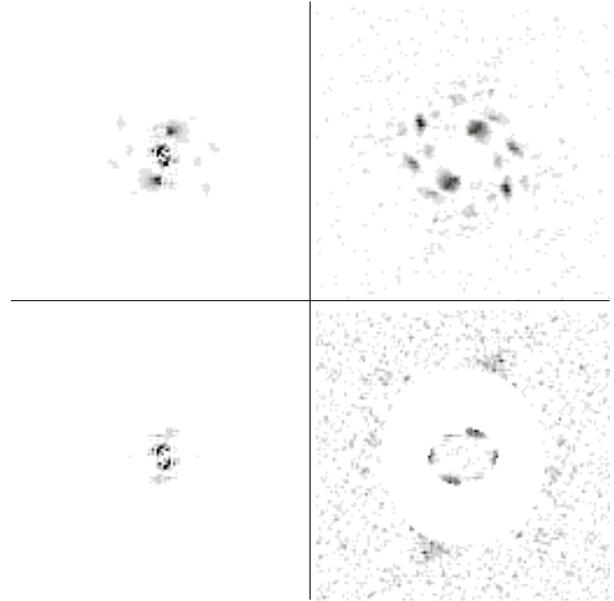,width=8cm}
\caption{\label{ChiPlot}These images show the contribution of each pixel to the symmetry $\chi^2$ for visit 11,
using the best-fit optical galaxy positions. The top-left image shows the per-pixel $\chi^2$ contributions when
no masking is used and no PSF error is used. The image on the top-right shows the contributions when no masking is
applied and a 7\% PSF error is used. The bottom-left image shows the contributions when spiral arms are masked, with 
no PSF error. The bottom-right image shows the $\chi^2$ contributions when the prominent spiral arms are
masked and a 7\% PSF error is used. All images are 128 pixels (3.2\arcsec) in both width and height.}
\end{figure}

In Figure \ref{ChiPlot} we show the per-pixel $\chi^2$ contributions between the two cases of no masking and
masked spiral arms for the optimum galaxy position, together with the contributions when no masking is used 
and zero PSF error is applied. With zero PSF error it is clear that the residuals from the 
subtraction of A and B dominate the $\chi^2$ measure. The effect of a non-zero PSF error is to suppress the 
A/B subtraction residuals and cause the spiral arms to dominate the $\chi^2$ measure, unless they are masked.
Because it is unclear which position (masked or unmasked) best represents the mass centre of the lensing galaxy, 
we report both masked and unmasked galaxy positions (and hence estimates for $H_0$) on equal terms.

\begin{figure*}
\psfig{figure=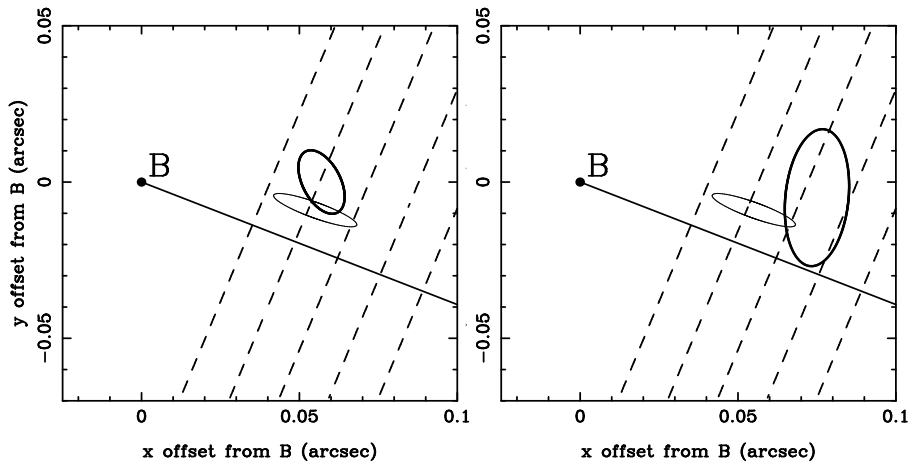,width=12cm}
\caption{\label{AllPlot}
The optical galaxy position compared to that determined by Wucknitz et al. (2004) using \lc. 
The error ellipses are 95\% confidence regions. The left-hand plot is for the case with no masking. 
The right-hand plot represents the case with masking of spiral arms. The position of B is marked, 
as is a line pointing towards the A component. The dotted lines are contours of $H_0$ in the strictly
isothermal case, and correspond (from right to left) to $H_0$ = (90,80,70,60,50)~\hunits. The positions
determined from optical data are shown as bold ellipses.}
\end{figure*}

Deriving errors on position from the individual visits is difficult because the symmetry $\chi^2$ increases very rapidly away from the minimum.
An error measure derived from the shape of the minimum for a single visit implies a spuriously high accuracy for the galaxy position.
It is likely that the number of degrees of freedom in the fit is over-estimated and that many pixels do not contribute any useful 
information to the fit statistic, since the drizzling process introduces correlations between neighbouring drizzled pixels. However, the scatter 
between positions derived from different visits is large. We therefore estimate errors on the galaxy position by taking ellipses that enclose 
68\% and 95\% of the measurements from all visits to define our $1\,\sigma$ and $2\,\sigma$ confidence levels. Figure \ref{AllPlot} shows the 
95\% confidence ellipses on the galaxy position for both sets of fits, as well as the position derived from \lc\ applied to VLA data by Wucknitz 
et al. (2004). 


The positions obtained from the symmetry fits were combined with the 
extra constraints available from the VLBI substructure described in Patnaik et al. (1995) and Kemball et al. (2001), which were 
used to constrain mass models by Wucknitz (2004). The optical galaxy position was combined with the models 
of Wucknitz (2004) by adding $\chi^2$ values for the galaxy position to the $\chi^2$ values from 
the lens models. However, the $\chi^2$ values taken from the symmetry fitting grid have too many degrees of freedom, and
so we assume that the optical position minimum is parabolic and form a new $\chi^2$ statistic based on our 68\% and 95\% 
confidence ellipses. We define the new $\chi^2$ statistic to have a value of 2.31 on our 68\% confidence ellipse, and a value
of 5.99 on the 95\% ellipse, and sum this statistic with that from the lens modelling. We emphasize that the scatter between 
visits dominates the random error budget for our measurement of $H_0$. 

Combining the VLBI and optical constraints shows that the best-fit galaxy position and the optical galaxy 
position are not coincident, as shown in Table \ref{AllTable}. The galaxy position shifts by up to 13~mas between the optical fit
and the optical+VLBI fit. The shapes of the confidence regions are also altered. However, the value of $H_0$ is not very sensitive 
to this mainly northerly shift. 

\begin{table*}
\begin{tabular}{lllccccccc}
Data used & Masking & Mass profile & \multicolumn{2}{c}{Position (mas)} & $H_0$ & $H_0$ & $\beta$ & Ellipticity\\
&&& $\Delta\alpha$ & $\Delta\delta$ & (isothermal) & (variable $\beta$) & & (of potential)\\
\\
Optical & None & - & $+$57 & $+$0 & 79$\pm$7 & 68$\pm$6 & 1.13$^{+0.07}_{-0.09}$ & 0.08$\pm$0.03 \\
Optical & Spiral & - & $+$75 & $-$5 & 66$\pm$9 & 56$^{+12}_{-15}$ & 1.16$\pm$0.19 & 0.05$\pm$0.04 \\
VLBI+Optical & None & Isothermal & $+$60 & $-$13 & 74$\pm$5 & - & - & 0.05$\pm$0.02 \\
VLBI+Optical & Spiral & Isothermal & $+$74 & $-$19 & 64$\pm$7 & - & - & 0.03$\pm$0.02 \\
VLBI+Optical & None & Variable & $+$60 & $-$12 & - & 70$\pm$5 & 1.05$\pm$0.03 & 0.04$\pm$0.02 \\
VLBI+Optical & Spiral & Variable & $+$74 & $-$18 & - & 61$\pm$7 & 1.05$\pm$0.04 & 0.04$\pm$0.02 \\
\end{tabular}

\caption{\label{AllTable}
Lens galaxy positions and the corresponding values of $H_0$ and the mass profile slope $\beta$. The optical positions are derived from
the ACS images only. The ``VLBI+Optical'' positions incorporate constraints from the \lc-based lens modelling of Wucknitz et al. (2004). 
The ``Mass profile'' column indicates what mass profile was assumed when combining the VLBI and optical constraints.
$H_0$ values are given in \hunits. Position offsets are referenced to image B, and RA offsets are given taking west as positive. All 
errors are quoted at 95\% confidence.}
\end{table*}

\section{Extraction of $\bmath{H_0}$}

The general relation between the time delay $\Delta t_{i,j}$ between the 
$i^{th}$ and $j^{th}$ images, the Hubble constant $H_0$ and the lens model, 
parametrised by the potential $\psi$, is given by

\begin{equation}
c \, \Delta t_{i,j} = \frac{1+\zd}{H_0}\frac{\dd \, \ds}{\dds}{\left(\phi_i - \phi_j\right)} \rtext{,}
\end{equation}

where $\zd$ is the redshift of the lens, $\dd$ and $\ds$ are the angular size distances to the
lens and source, respectively, $\dds$ is the angular size distance to the
source measured from the lens, and $\phi_i$ is the scaled time delay at the position of the 
$i^{th}$ image ($\btheta_i$),

\begin{equation}
\phi_i = \frac12 |\nabla\psi(\btheta_i)|^2-\psi(\btheta_i) \rtext{.}
\end{equation}

The angular size distances are normalized in these equations, since they do not include factors of $H_0$.
For general isothermal models without external shear the relation becomes particularly simple and can be 
written as a function of the image positions alone, without explicitly using any lens model parameters 
(Witt et al. 2000):

\begin{equation}
\phi_i = \frac12 |\btheta_i - \btheta_0|^2
\end{equation}
Here $\vc \btheta_0$ is the position of the centre of the lens.
External shear $\gamma$ changes $\phi_i$ by a factor between $1\pm\gamma$
depending on the relative direction, typically resulting in similar
factors for the value deduced for $H_0$. A general analysis for power-law models with
external shear can be found in Wucknitz (2002).

Using the recipe described in previous sections our lens position translates to a
Hubble constant of H$_0=79\pm7$~\hunits in the shearless isothermal
case\footnote{A concordance cosmological model with $\Omega=0.3$ and
$\lambda=0.7$ and a homogeneous matter distribution is used for the
calculation of all distances in this paper} without masking, and to $66\pm9$ with masking. 

Estimates of external shear and convergence from nearby field galaxies and large scale
structure are of the order 2~per~cent (Leh\'ar et al. 2000) and would 
affect the result only to the same relative amount, sufficiently below our
current error estimate to allow us to neglect these effects.

\begin{figure*}
\begin{tabular}{cc}
\psfig{figure=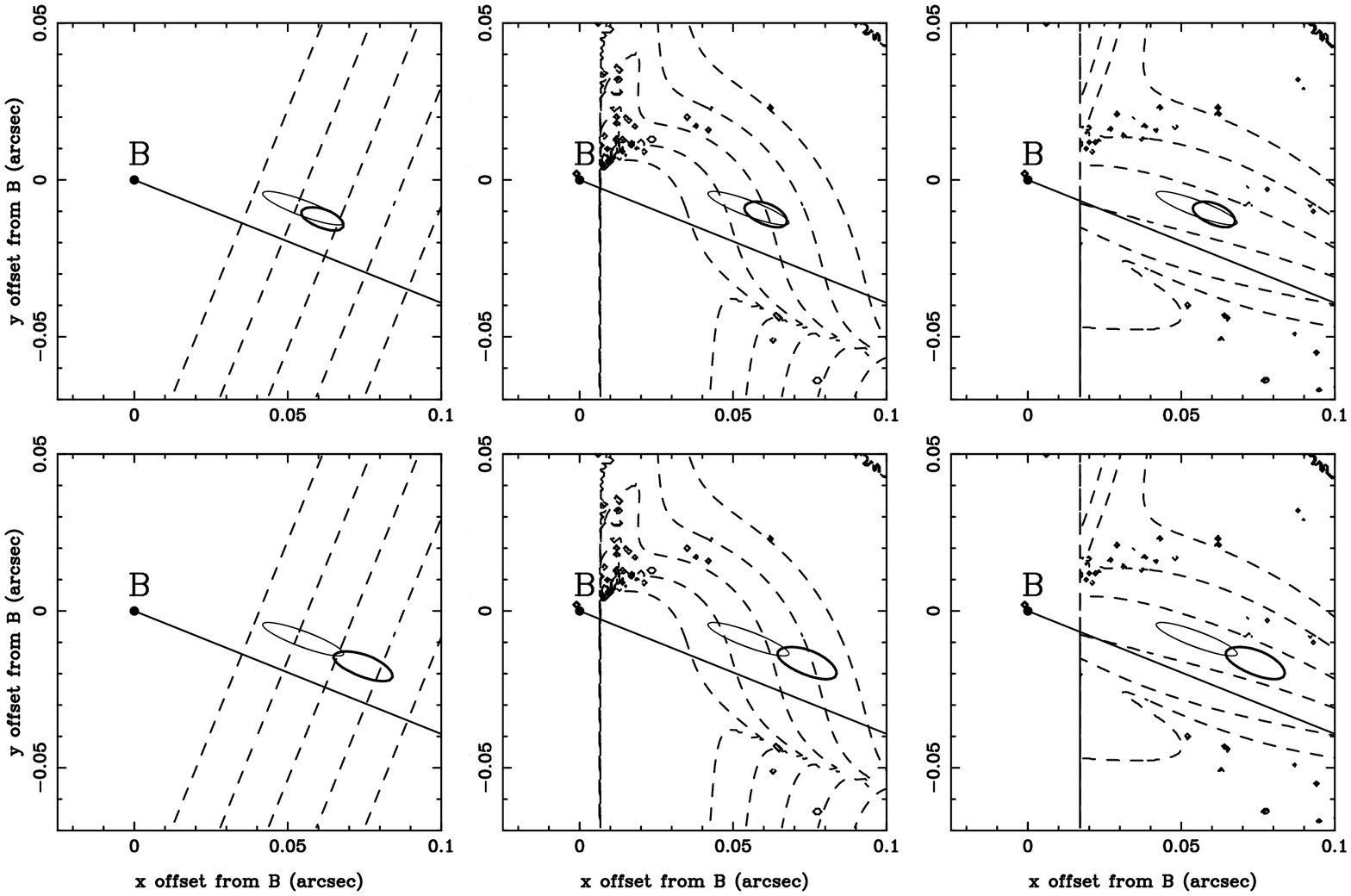,width=17.5cm}
\end{tabular}
\caption{\label{TypePlot} The 95\% confidence regions for the cases of no masking (top row) and masked spiral
arms (bottom row). The plots in the left-hand column show the galaxy position confidence regions and contours of $H_0$ calculated 
for isothermal models. The plots of the central column show the confidence regions superimposed over contours of $H_0$ 
calculated whilst allowing the logarithmic mass slope $\beta$ to vary. This reduces $H_0$ slightly compared 
to the fixed $\beta=1$ (isothermal) case. The plots in the right-hand column show the same confidence regions superimposed 
over contours of $\beta$. The confidence regions take both our optical position and the VLBI constraints used 
by Wucknitz et al. (2004) into account. The 95\% confidence error ellipse of Wucknitz et al. is also shown.
Contours of $H_0$ are again at (90,80,70,60,50)\hunits reading from left to right across an image, and 
contours of $\beta$ are (0.8,0.9,1.0,1.1,1.2,1.3) reading from bottom to top. The optical+VLBI error ellipses are shown in bold
relative to the \lc\ ellipse of Wucknitz et al.}
\end{figure*}

The value of the Hubble constant we derive depends on the slope of the
mass distribution of the lensing galaxy. In Figure \ref{TypePlot} we show the permitted values of
the Hubble constant for different models -- isothermal and with a variable
$\beta$ in an elliptical potential model -- plotted against measured galaxy
position. The elliptical power-law potential is parametrised
as follows:
\begin{gather}
\psi(\btheta) = \frac{\theta\sub E^{2-\beta}}{\beta} \, r_\epsilon^\beta (\btheta)
\rtext{,} \\ 
r_\epsilon^2 = \frac{\theta_x^2}{(1+\epsilon)^2} + \frac{\theta_y^2}{(1-\epsilon)^2}
\rtext{,} \\
\btheta = (\theta_x, \theta_y)
\rtext{,}
\end{gather}
where $\theta_E$ is the Einstein radius of the model, $\beta$ is the power-law index of the 
potential's profile and $\epsilon$ is the ellipticity of the potential. For details of our modelling procedure the reader is referred to
Wucknitz et al. (2004). It is evident that the preferred value of the
Hubble constant is somewhat reduced compared to what is obtained by
forcing the mass distribution to be isothermal. We also show
contours of the radial power law $\beta$ plotted against galaxy
position. The optical lens position gives $\beta=$1.13$^{+0.07}_{-0.09}$ ($2\,\sigma$). 

As discussed before, B0218+357 has the advantage of clear substructure in
the two images which can be mapped with VLBI. The VLBI data can be used
independently to derive the slope of the mass profile of the lens (Wucknitz
et al., 2004). Biggs et al. (2003) and Wucknitz et al. find a value of $\beta$=1.04$\pm$0.02.
Combining the VLBI constraints with the optical lens position gives a value of $\beta$=1.05$\pm$0.03 (95\% confidence).
We therefore adopt this value for the mass profile's logarithmic slope and obtain a Hubble constant of 70$\pm$5~\hunits (95\% confidence)
for the case with no masking and 61$\pm$7~\hunits (95\% confidence) when the spiral arms are masked.
The ellipticity of the potential is small (about 0.04) in each case, but the ellipticity of the mass distribution
will be about three times this, 0.12. The lens galaxy could be more inclined than it appears, or there could be
a bar or similar feature present.

\section{Conclusions}

We have analysed the deepest optical image yet taken of B0218+357 
to measure the position of the lens galaxy. We find that simple 
subtraction of a parametric galaxy model and two point sources 
is insufficient to constrain the galaxy position, and we confirm 
earlier suggestions that the image separation in the optical is lower
than that in the radio, most probably due to significant extinction around 
image A. Taking advantage of the symmetric appearance of the lens, we have defined 
the centre as that point about which the residuals (after subtraction of 
A and B) are most symmetric. To account for artifacts in our empirical
PSF model we have introduced an extra noise term. We have also masked off
the most prominent spiral arms to test the effect on $H_0$. We
find that the lens galaxy position is 57$\pm$4~mas west and 0$\pm$6~mas south
of image B when no masking is applied. Combined with the results of Wucknitz et al. (2004) 
this leads to a value for $H_0$ of 70$\pm$5~\hunits (95\% confidence). When the most obvious spiral arms 
are masked out, we find an optical galaxy position of 75$\pm$6~mas west and $-$5$\pm$13~mas south from image B. 
This results in a value for $H_0$ of 61$\pm$7~\hunits (95\% confidence) when combined with VLBI constraints.

Further work on this lens will involve increased use of \lc\ to further limit the power law exponent $\beta$ using VLBI constraints. 
Observations have also been made using the VLA with the Pie Town VLBA antenna, which
together with VLBI will further improve the lens model for this system.

\section*{Acknowledgments}

The {\it Hubble Space Telescope} is operated at the Space Telescope Science
Institute by Associated Universities for Research in Astronomy Inc.
under NASA grant NAS5-26555. We would like to thank the ACS team, 
especially Warren Hack, Max Mutchler, Anton Koekemoer and Nadezhda 
Dencheva for help and advice on the data reduction. We would also like to thank 
the referee, Chris Fassnacht, whose comments greatly improved the substance of the paper.

TY acknowledges a PPARC research studentship. OW was supported by the BMBF/DLR Verbundforschung 
under grant 50~OR~0208. This research has made use of the NASA/IPAC 
Extragalactic Database. {\sc iraf} is distributed by the National 
Optical Astronomy Observatories, which are operated by AURA, Inc., 
under cooperative agreement with the National Science Foundation.

\section*{References}

\iftrue
\newcommand{\mnras}{\mbox{MNRAS}}
\newcommand{\apj}{\mbox{ApJ}}
\newcommand{\aj}{\mbox{AJ}}
\newcommand{\aaps}{\mbox{A\&AS}}
\newcommand{\aap}{\mbox{A\&A}}
\bibliographystyle{mn2e}
\bibliography{paper}
\fi

\normalsize

\noindent Barkana R., 1997, ApJ, 489, 21

\noindent Barkana R., Leh\'ar J., Falco E.E., Grogin N.A., Keeton C.R., Shapiro I.I.,
1999, ApJ, 520, 479

\noindent Bernstein G., Fischer P., 1999, AJ, 118, 14

\noindent Biggs A.D., Browne I.W.A., Helbig P., Koopmans L.V.E., Wilkinson P.N., Perley R.A.,
 1999, MNRAS, 304, 349

\noindent Biggs A.D., Xanthopoulos E., Browne I.W.A., Koopmans L.V.E., Fassnacht C.D.,
 2000, MNRAS, 318, 73

\noindent Biggs A.D., Browne I.W.A., Muxlow T.W.B., Wilkinson P.N.,
 2001, MNRAS, 322, 821

\noindent Biggs A.D., Wucknitz O., Porcas R.W., Browne I.W.A.,
Jackson N.J., Mao S., Wilkinson P.N., 2003, MNRAS, 338, 599


\noindent Browne I.W.A., Patnaik A.R., Walsh D., Wilkinson P.N.,
 1993, MNRAS, 263, L32

\noindent Burud I., Hjorth J., Jaunsen A.O., Andersen M.I., 
Korhonen H., Clasen J.W., Pelt J., Pijpers F.P., Magain P., Ostensen R.,
 2000, ApJ, 544, 117

\noindent Burud I., Hjorth J., Courbin F., Cohen J.G., Magain P.,
Jaunsen A.O., Kaas A.A., Faure C., Letawe G., 2002a, A\&A, 391, 481

\noindent Burud I., et al., 2002b, A\&A, 383, 71

\noindent Carilli C.L., Rupen M.P., Yanny B., 1993, ApJL, 412, L59

\noindent Clampin M., et al., 2000, SPIE, 4013, 344

\noindent Cohen, A.S., Hewitt, J.N., Moore, C.B., Haarsma, D.B., 2000, ApJ, 545, 578

\noindent Cohen J.G., Lawrence C.R., Blandford R.D.,
2003, ApJ, 583, 67

\noindent Cohn J.D., Kochanek C.S., McLeod B.A., Keeton C.R., 2001, ApJ, 554, 1216

\noindent Combes F., Wiklind T., 1997, ApJ, 486, L79

\noindent Ellithorpe J.D., Kochanek C.S., Hewitt J.N.,
 1996, ApJ, 464, 556

\noindent Falco E.E., Gorenstein M.V., Shapiro I.I.,
 1985, ApJL, 289, L1

\noindent Fassnacht C.D., Pearson T.J., Readhead A.C.S., Browne I.W.A., Koopmans L.V.E., Myers S.T., Wilkinson P.N.,
 1999, ApJ, 527, 498

\noindent Fassnacht C.D., Xanthopoulos E., Koopmans L.V.E., Rusin D.,
 2002, ApJ, 581, 823

\noindent Freedman W.L., Madore B.F., Gibson B.K., Ferrarese L., Kelson D.D., Sakai S., Mould J.R., Kennicutt R.C., Jr., Ford H.C., Graham J.A., et al.,
 2001, ApJ, 553, 47

\noindent Fruchter A.S., Hook R.N., 2002, PASP, 114, 144

\noindent Gorenstein M.V., Shapiro I.I., Falco E.E., 1988,
ApJ, 327, 693

\noindent Grundahl F., Hjorth J., 1995, MNRAS, 275, L67

\noindent Hack W.J., 2002, in Bohlender D., Durand D., Handley T.H., eds, 
ASP Conf. Ser. Vol. 281, Astronomical Data Analysis Software and Systems 
XI. Astron. Soc. Pac., San Francisco, p.197.

\noindent Hjorth J., 1997, Helbig P., Jackson N., eds, Proc. Golden Lenses,
Hubble's Constant and Galaxies at High Redshift Workshop. Jodrell Bank Observatory,
Cheshire.\footnote{Proceedings available from http://www.jb.man.ac.uk/research/gravlens/workshop1/prcdngs.html}

\noindent Hjorth J., et al., 2002, ApJ, 572, L11


\noindent Jackson N., Xanthopoulos E., Browne I.W.A., 2000, MNRAS, 311, 389

\noindent Keeton C.R., et al., 2000, ApJ, 542, 74

\noindent Kemball A.J., Patnaik A.R., Porcas R.W., 2001, ApJ, 562, 649

\noindent Kochanek C.S., Keeton C.R., McLeod B.A., 2001, ApJ, 547, 50

\noindent Kochanek C.S., Narayan R., 1992, ApJ, 401, 461

\noindent Kochanek C.S., 2002, ApJ, 578, 25

\noindent Kochanek C.S., 2003, ApJ, 583, 49

\noindent Kochanek C.S., Schechter P., 2004, in ``Measuring and Modelling the Universe'',
Carnegie Obs. Centennial Symposium, CUP, ed. W. Freedman, p. 117

\noindent Koekemoer A.M., Fruchter A.S., Hook R.N., Hack W., 2002,
in Arriba S., ed., The 2002 HST Calibration Workshop : Hubble after 
the Installation of the ACS and the NICMOS Cooling System. 
Space Telescope Science Institute, Baltimore, MD, p. 339

\noindent Koopmans L.V.E., de Bruyn A.G., Xanthopoulos E., Fassnacht C.D., 2000,
A\&A, 356, 391

\noindent Koopmans L.V.E., Fassnacht C.D., 1999, ApJ, 527, 513

\noindent Koopmans L.V.E., Treu T., 2003, ApJ, 583, 606

\noindent Krist J., 1995, in Shaw R.A., Payne H.E., Hayes J.J.E., eds, ASP Conf. Ser.
Vol. 77, Astronomical Data Analysis Software and Systems IV. Astron. Soc. Pac., 
San Francisco, p. 349

\noindent Kundic T., Hogg D.W., Blandford R.D., Cohen J.G., Lubin L.M., Larkin J.E.,
 1997, AJ, 114, 2276

\noindent Landolt A.U., 1992, AJ, 104, 340

\noindent Leh\'ar J., Falco E.E., Kochanek C.S., McLeod B.A., Mu\~noz J.A., Impey C., Rix H., Keeton C.R., Peng C.Y.,
 2000, ApJ, 536, 584

\noindent Lovell J.E.J., Jauncey D.L., Reynolds J.E., Wieringa M.H., King E.A.,
Tzioumis A.K., McCulloch P.M., Edwards P.G., 1998, ApJ, 508, L51

\noindent Mack J., et al., 2003, ACS Data Handbook, Version 2.0 (Baltimore:STScI)

\noindent McLure R.J., Kukula M.J., Dunlop J.S., Baum S.A., O'Dea C.P., Hughes D.H.,
 1999, MNRAS, 308, 377

\noindent Menten K.M., Reid M.J., 1996, ApJL, 465, L99

\noindent Merline W., Howell S.B., 1995, Expt. Astron., 6, 163

\noindent Mould, J.R., et al., 2000, ApJ, 529, 786

\noindent Mu\~noz J.A., Kochanek C.S., Keeton C.R.,
 2001, ApJ, 558, 657

\noindent Mutchler M., Cox C., 2001, Instrument Science Report 
ACS 2001-07 (Baltimore: STScI)

\noindent Ofek E.O., Maoz D., 2003, ApJ, 594, 101

\noindent Oscoz A., Alcalde D., SerraRicart M., Mediavilla E., Abajas C., Barrena R., Licandro J., Motta V., Mu\~noz J.A.,
 2001, ApJ, 552, 81

\noindent Patnaik A.R., Browne I.W.A., Wilkinson P.N., Wrobel J.M., 
1992, MNRAS, 254, 655

\noindent Patnaik A.R., Browne I.W.A., King L.J., Muxlow T.W.B., Walsh D.,
Wilkinson P.N., 1993, MNRAS, 261, 435

\noindent Patnaik A.R., Porcas R.W., Browne I.W.A., 1995, MNRAS, 274, L5

\noindent Patnaik A.R., Narasimha D., 2001, MNRAS, 326, 1403

\noindent Pavlovsky C., et al., 2002, ACS Instrument Handbook, Version 3.0
(Baltimore: STScI).

\noindent Refsdal S., 1964, MNRAS, 128, 307

\noindent Rusin D., Ma C., 2001, ApJL, 549, L33

\noindent Saha P., Williams L.L.R., 2001, AJ, 122, 585

\noindent Schechter P.L., 2001, in Brainerd T.G.,
Kochanek C.S., eds, ASP Conf. Ser. Vol. 237, Gravitational
Lensing: Recent Progress and Future Goals. 
Astron. Soc. Pac., San Francisco, p. 427

\noindent Schechter P.L., Bailyn C.D., Barr R., Barvainis R., Becker C.M., Bernstein G.M., Blakeslee J.P., Bus S.J., Dressler A., Falco E.E., et al.,
 1997, ApJL, 475, L85

\noindent Stickel M., Kuehr H., 1993, A\&AS, 101, 521

\noindent Sykes C.M., et al., 1998, MNRAS, 301, 310

\noindent Walsh D., Carswell R.F., Weymann R.J., 1979, Nature, 279, 381

\noindent Wiklind T., Combes F., 1995, A\&A, 299, 382

\noindent Williams R.E., et al., 1996, AJ, 112, 1335

\noindent Williams L.L.R., Saha P., 2000, AJ, 119, 439

\noindent Witt H.J., Mao S., Keeton C.R., 2000, ApJ, 544, 98

\noindent Wucknitz O., 2002, MNRAS, 332, 951

\noindent Wucknitz O., 2004, MNRAS, 349, 1

\noindent Wucknitz O., Biggs A.D., Browne I.W.A., 2004, MNRAS, 349, 14

\label{lastpage}
\end{document}